

\input phyzzx.tex

\overfullrule=0pt

\let\refmark=\NPrefmark 
\def\define#1#2\par{\def#1{\Ref#1{#2}\edef#1{\noexpand\refmark{#1}}}}
\def\con#1#2\noc{\let\?=\Ref\let\<=\refmark\let\Ref=\REFS
         \let\refmark=\undefined#1\let\Ref=\REFSCON#2
         \let\Ref=\?\let\refmark=\<\refsend}

\define\RZWB
B. Zwiebach, preprint IASSNS-HEP-92/41 (hepth/9206084) and references
therein.

\define\RGIDST
S. Giddings and A. Strominger, Phys. Rev. Lett. {\bf 67} (1991) 2930.

\define\RSAROJA
R. Saroja and A. Sen, Phys. Lett. {\bf B286} (1992) 256.

\define\RFMS
D. Friedan, E. Martinec and S. Shenker, Nucl. Phys. {\bf B271} (1986) 93.

\define\RNEWST
D. Waldran, preprint EFI-92-43 (hepth/9210031).

\define\RSTRING
A. Sen, preprint TIFR-TH-92-39 (hepth/9206016) (to appear in Nucl.
Phys. B).

\define\RDGHR
A. Dabholkar, G. Gibbons, J. Harvey and F.R. Ruiz, Nucl. Phys. {\bf B340}
(1990) 33;
A. Dabholkar and J. Harvey, Phys. Rev. Lett. {\bf 63} (1989) 719.

\define\RGIB
G. Gibbons, Nucl. Phys. {\bf B207} (1982) 337;
G. Gibbons and K. Maeda, Nucl. Phys. {\bf B298} (1988) 741.

\define\RODDMORE
M. Rocek and E. Verlinde, preprint IASSNS-HEP-91-68;
A.A. Tseytlin, Mod. Phys. Lett. A6 (1991) 1721;
in Proc.  First Int. A.D. Sakharov Conference on Physics,
ed. by L.V. Keldysh et al. (Nova Science Pub., Commack, NY, 1991);
A.A. Tseytlin and C. Vafa, Nucl. Phys. B372 (1992) 443;
A. Giveon, Mod. Phys. Lett. A6 (1991) 2843;
R. Dijkgraaf, E. Verlinde and H. Verlinde, Nucl. Phys. B371 (1992) 269;
E. Smith and J. Polchinski, Phys. Lett. B263 (1991) 59;
T. Kugo and B. Zwiebach, Prog. Theor. Phys. 87 (1992) 801.

\define\RHOHO
J. Horne and G. Horowitz, preprint UCSBTH-92-11 (hepth/9203083).

\define\RHOST
D. Garfinkle, G. Horowitz and A. Strominger, Phys. Rev. {\bf D43} (1991)
3140;
G. Horowitz and A. Strominger, Nucl. Phys. {\bf B360} (1991) 197.

\define\RGAZU
M. Gaillard and B. Zumino, Nucl. Phys. {\bf B193} (1981) 221.

\define\RWIT
E. Witten, Phys. Rev. {\bf D44} (1991) 314;
C. Nappi and E. Witten, preprint IASSNS-HEP-92/38 (hepth/9206078)
and references therein.

\define\RTWOD
G. Mandal, A. Sengupta and S. Wadia, Mod. Phys. Lett. {\bf A6} (1991)
1685.

\define\RODDNEW
J. Horne, G. Horowitz and A. Steif, Phys. Rev. Lett. {\bf 68}
(1992) 568;
P. Horava, preprint EFI-91-57;
J. Panvel, Phys. Lett. {\bf B284} (1992) 50.

\define\RKUM
S. Kar, S. Khastagir and A. Kumar, Mod. Phys. Lett. {\bf A7} (1992) 1545;
S. Kar and A. Kumar, preprint IP-BBSR-92-18 (hepth/9204011);
J. Maharana, preprint CALT-68-1781 (hepth/9205016);
S. Khastagir and J. Maharana, preprint IP-BBSR-92-38 (hepth/9206017);
A. Kumar, preprint CERN-TH.6530/92 (hepth/9206051).

\define\RODD
S. Ferrara, J. Scherk and B. Zumino, Nucl. Phys. {\bf B121} (1977) 393;
E. Cremmer, J. Scherk and S. Ferrara, Phys. Lett. {\bf B68} (1977) 234;
{\bf B74} (1978) 61;
E. Cremmer and J. Scherk, Nucl. Phys. {\bf B127} (1977) 259;
E. Cremmer and B. Julia, Nucl. Phys.{\bf B159} (1979) 141;
M. De Roo, Nucl. Phys. {\bf B255} (1985) 515; Phys. Lett. {\bf B156}
(1985) 331;
E. Bergshoef, I.G. Koh and E. Sezgin, Phys. Lett. {\bf B155} (1985) 71;
M. De Roo and P. Wagemans, Nucl. Phys. {\bf B262} (1985) 646;
L. Castellani, A. Ceresole, S. Ferrara, R. D'Auria, P. Fre and E. Maina,
Nucl. Phys. {\bf B268} (1986) 317; Phys. Lett. {\bf B161} (1985) 91;
S. Cecotti, S. Ferrara and L. Girardello, Nucl. Phys. {\bf B308} (1988)
436;
M. Duff, Nucl. Phys. {\bf B335} (1990) 610.

\define\RVENEZIANO
G. Veneziano, Phys. Lett. {\bf B265} (1991) 287;
K. Meissner and G. Veneziano, Phys. Lett. {\bf B267} (1991) 33; Mod. Phys.
Lett. {\bf A6} (1991) 3397;
M. Gasperini, J. Maharana and G. Veneziano, Phys. Lett. {\bf B272} (1991)
277; preprint CERN-TH.6634/92 (hepth/9209052);
M. Gasperini and G. Veneziano, Phys. Lett. {\bf B277} (1992) 256.

\define\RGIVEON
A. Giveon and M. Rocek, Nucl. Phys. {\bf B380} (1992) 128;
A. Giveon and A. Pasquinucci, preprint IASSNS-HEP-92/35
(hepth/9208076).

\define\RSEN
A. Sen, Phys. Lett. {\bf B271} (1991) 295; {\bf 274} (1991) 34.

\define\RFAWAD
S.F. Hassan and A. Sen, Nucl. Phys. {\bf B375} (1992) 103.

\define\RROTBLA
A. Sen, Phys. Rev. Lett. {\bf 69} (1992) 1006.

\define\RSDUAL
A. Sen, preprint TIFR/TH/92-41 (hepth/9207053).

\define\RORTIN
T. Ortin, preprint SU-ITP-92-24 (hepth/9208078).

\define\REARBL
R. C. Myers and M. Perry, Ann. Phys. {\bf 172} (1986) 304;
R.C. Myers, Nucl. Phys. {\bf B289} (1987) 701;
P. Mazur, Gen. Rel. and Grav. {\bf 19} (1987) 1173;
C. G. Callan, R. C. Myers and M. Perry, Nucl. Phys. {\bf B311} (1988) 673;
H.J. de Vega and N. Sanchez, Nucl. Phys. {\bf B309} (1988) 552;
I. Ichinose and H. Yamazaki, Mod. Phys. Lett. {\bf A4} (1989) 1509;
Class. Quantum Grav. {\bf 9} (1992) 257;
M. Fabbrichesi, R. Iengo and K. Roland, preprint SISSA/ISAS 52-92-EP.

\define\RKALLOSH
R. Kallosh, Phys. Lett. {\bf B282} (1992) 80;
R. Kallosh, A. Linde, T. Ortin, A. Peet and A. Van Proeyen, preprint
SU-ITP-92-13 (hepth/9205027).

\define\RKHURI
R. Khuri, preprints CTP-TAMU-34-92 (hepth/9205051); CTP-TAMU-35-92\break
\noindent (hepth/9205081);
CTP-TAMU-38-92 (hepth/9205091); CTP-TAMU-44-92 (hepth/9205108).

\define\RMAHSCH
J. Maharana and J. Schwarz, preprint CALT-68-1790 (hepth/9207016);
J. Schwarz, preprint CALT-68-1815 (hepth/9209125).

\define\RTSW
A. Shapere, S. Trivedi and F. Wilczek, Mod. Phys. Lett. {\bf A6}
(1991) 2677.

\define\RCAMP
B. Campbell, M. Duncan, N. Kaloper and K. Olive, Phys. Lett. {\bf B251}
(1990) 34;
B. Campbell, N. Kaloper and K. Olive, Phys. Lett. {\bf B263} (1991) 364;
{\bf B285} (1991) 199.

\define\RSTRSOL
A. Strominger, Nucl. Phys. {\bf B343} (1990) 167;
C. G. Callan, J. Harvey and A. Strominger, Nucl. Phys. {\bf B359} (1991)
611; Nucl. Phys. {\bf B367} (1991) 60; preprint EFI-91-66.

\define\RDUFFLU
M. Duff and J. Lu, Phys. Rev. Lett. {\bf 66} (1991) 1402;
Class. Quantum Grav. {\bf 9} (1991) 1;
Nucl. Phys. {\bf B354} (1991) 141.

\def\R{{\cal R}}
\def\Chi{\chi}
\def\eps{\epsilon}
\def\den{(\rho^2 +a^2\cos^2\theta +2m\rho\sinh^2{\alpha\over 2})}
\def\p{\partial}
\def\tF{\tilde F}
\def\bl{\bar\lambda}
\def\l{\langle}
\def\r{\rangle}
\def\ch{{\cal H}}
\def\cg{{\cal G}}
\def\co{{\cal O}}
\def\ck{{\cal K}}
\def\cm{{\cal M}}
\def\odd{O(d)\times O(d)}

{}~\hfill\vbox{\hbox{TIFR-TH-92-57}\hbox{hepth@xxx/9210050}\hbox{October,
1992}}\break

\title{BLACK HOLES AND SOLITONS IN STRING THEORY}

\author{Ashoke Sen}

\address{Tata Institute of Fundamental Research, Homi Bhabha Road, Bombay
400005, India}
\centerline{e-mail address: sen@tifrvax.bitnet}

\abstract
In this talk, I discuss a general method for
constructing classical
solutions of the equations of motion arising in the effective low energy
string theory,
and discuss specific applications of this method.

\noindent (Based on talks given at the Johns Hopkins Workshop held at
Goteborg, June
8-10, 1992, and ICTP Summer Workshop held at Trieste, July 2-3, 1992)

\chapter{Introduction}

In recent years there has been much interest in the construction of
classical soliton\con\RDGHR\RSTRSOL\RDUFFLU\RKHURI\noc\ and black
hole\con\REARBL\RGIB\RHOST\RGIDST\RKALLOSH\noc\
solutions in string theory, including those in
low\RTWOD\ dimensions,  and those which correspond to solvable
conformal field theories\RWIT.
There are several motivations for studying classical solutions in string
theory; I shall only quote two of them here.
First of all, one must keep in  mind that the full spectrum of string
theory should include soliton like solutions that might be present in
string  theory, and various non-perturbative symmetries in string theory
may become manifest
only after including all the soliton states in the spectrum.
Secondly, since string theory is expected to provide us with a finite,
consistent theory of quantum gravity, various puzzles involving black
hole evaporation might be resolved by studying them in the
context of string theory.
This analysis would require construction of black hole solutions in this
theory.

In this talk I shall discuss a general method for
generating new classical solutions from known
solutions\con\RVENEZIANO\RSEN\RFAWAD\RODDNEW\RKUM\RTSW\RSDUAL\noc.
The  general idea  is as
follows.
Let $\phi^i$ be the fields and $S(\phi^i)$ be the action of a classical
field theory, and $G$ be the group of symmetries of the action $S$.
In that case $G$ is also the group of symmetries of the equations of
motion, and if $\phi_i^{(cl)}$ is a solution of the equations of motion,
then, $\forall g\in G$, $g\phi_i^{(cl)}$ is also a solution of the
equations of motion.
However, $\phi_i^{(cl)}$ and $g\phi_i^{(cl)}$ are equivalent solutions,
in the sense that no physical experiment can distinguish between these
two solutions.

In a generic situation, if we restrict $\phi_i$ to a special class $K$ of
field configurations (e.g. field configurations that are independent of
$d$ of the dimensions) and write down an expression for the action, the
symmetry of the corresponding action will be a subgroup $H$ of $G$
consisting
of those elements of $G$ which keep $\phi_i$ inside the class $K$.
However, in some cases  the relevant symmetry group $\cg$ may be bigger than
$H$, consisting of elements outside the original symmetry group $G$.
If $\tilde g$ denotes such an element of $\cg$, and $\phi_i^{(cl)}$
denotes a  solution of the equations of motion belonging to the
class $K$, then $\tilde g\phi_i^{(cl)}$ denotes a new  solution
of the equations of motion of the full theory.
In this case, however this represents a physically inequivalent solution,
since $\tilde g$ is not an element of the symmetry group $G$ of the theory.

Thus if string theory happens to possess such extended symmetries when
we restrict field configurations to some special classes, we might use
them to generate new classical solutions from known ones.
I shall show that there are at least two classes of such symmetries in
string theory.
Both these classes of symmetries were studied in the context of
supergravity theories in the past\RGAZU\RODD.
The first class of
symmetries\RVENEZIANO-\RKUM\con\RGIVEON\RMAHSCH\RODDMORE\noc\ allows
us, in particular, to construct
charged solutions starting from charge neutral
ones.
The second class of symmetries allows us to construct solutions carrying
magnetic charge starting from solutions carrying electric
charge\RTSW\RSDUAL\RORTIN.

This review will be organised as follows.
In sect.2 I shall explain the origin of the first class of symmetries
using the language of string field theory and show that they are valid at
the string tree level to all orders in $\alpha'$.
In sect.3 I shall see how this symmetry manifests itself in the low
energy effective field theory.
Sects.4 and 5 will be devoted to specific applications of this
transformation towards constructing new classical solutions of string
theory.
In sect.4 I shall discuss the construction of rotating charged black hole
solutions in string theory in four dimensions.
Sect.5 will be devoted to the construction of static classical solutions
that represent fundamental heterotic string carrying electric charge and
electric current.
In sect.6 I shall discuss the second class of symmetries  $-$ the
electric-magnetic duality transformation $-$ of the equations of motion
of string theory in four dimensions; and show how, using this
transformation, one can construct rotating black hole solutions in string
theory carrying both, electric and magnetic charges.

As a guide to the reader, I would suggest that those who are completely
unfamiliar with the concept of string field theory may skip section 2; the
rest of the article can be read without this section.
At the same time, I would like to emphasize that knowledge of string field
theory required for reading section 2 is minimal.

\chapter{String Field Theory Argument}

I shall first briefly review the formulation of string field
theory\RZWB.
For simplicity I shall consider the case of closed bosonic
string
field theory, although the case of superstring and heterotic string
field theory may be dealt with in the same way.
Let $\ch$ denote the Hilbert space of the combined matter-ghost conformal
field theory in two dimensions describing first quantized string theory.
The string field can be identified to a state $|\Chi\r$ in $\ch$ of ghost
number 2 (with
the convention that the SL(2,C) invariant vacuum has ghost number 0), and
annihilated by $L_0-\bar L_0$ and $b_0-\bar b_0$.
If $\{|\Chi_r\r\}$ is a basis of such states in $\ch$, then we may express
$|\Chi\r$ as $\sum_r \psi_r|\Chi_r\r$.
The $\psi_r$'s are the dynamical variables in string theory (the component
fields).
With this convention, the kinetic term in string theory is proportional to,
$$
\l\Chi|Q_B c_0^-|\Chi\r = \psi_r\psi_s\l\Chi_r|Q_B c_0^-|\Chi_s\r
\eqn\esftone
$$
where $Q_B$ is the BRST charge of the first quantised string theory, and
$c_0^-=(c_0-\bar c_0)/\sqrt 2$.
On the other hand, the interaction terms are given by,
$$
\sum_{n= 3}^\infty A_{r_1\ldots r_n}\psi_{r_1}\ldots \psi_{r_n}
\eqn\esfttwoa
$$
where $A_{r_1\ldots r_n}$ are coefficients calculated in terms of
correlation functions of certain conformal transform of the field
operators $\Chi_{r_1}, \ldots \Chi_{r_n}$ and ghost fields in the two
dimensional conformal field theory.
The specific form of $A_{r_1,\ldots r_n}$ will not be required for our
analysis.

Let us now consider the situation where the original background around
which we are formulating our string field theory is a free field theory of
$D$ scalar fields, together with another conformal field theory of central
charge $26-D$.
Restricting to field configurations independent of $d$ of the $D$
directions would amount to choosing the string field configuration
$|\tilde\Chi\r$ which carry zero momentum in these $d$ directions
$X^{D-d+1},\ldots X^D$.
Let $\{|\tilde\Chi_\alpha\r\}$ denote the subset of the basis vectors of
$\{|\Chi_r\r\}$ which carry zero momentum in these $d$ directions.
Then $|\tilde\Chi\r$ may be expanded as,
$$
|\tilde\Chi\r = \sum_\alpha \tilde\psi_\alpha |\tilde\Chi_\alpha\r
\eqn\esfttwo
$$
Thus the kinetic term of the restricted action will be given by,
$$
\tilde\psi_\alpha \tilde\psi_\beta \l\tilde\Chi_\alpha |Q_B
c_0^-|\tilde\Chi_\beta\r
\eqn\esftthree
$$
and the interaction terms are given by,
$$
\sum_{n=3}^\infty \tilde A_{\alpha_1\ldots\alpha_n}\tilde\psi_{\alpha_1}
\ldots\tilde\psi_{\alpha_n}
\eqn\esftfour
$$
The coefficients $\tilde A_{\alpha_1\ldots \alpha_n}$ are now computed in
terms of correlation functions of (conformal transforms of) the fields
$\tilde\Chi_{\alpha_1}, \ldots \tilde\Chi_{\alpha_n}$ and ghost fields in
the two
dimensional confomal field theory.
Since the fields $\tilde\Chi_{\alpha_i}$ carry zero momentum in the $d$
directions, part of the correlation function involving the fields
$X^{D-d+1},\ldots X^D$ factorises completely into holomorphic and
antiholomorphic parts.
Each of these parts must be separately invariant under the $d$ dimensional
Lorentz transformations.\foot{To see
this, let
us consider a correlation function of the form\break
\noindent $\l\p X^\mu(z,\bar z)\bar\p
X^\nu(z, \bar z) \p X^\tau(w,\bar w) \bar \p X^\sigma(w,\bar w)\r$.
This correlation function is proportional to\break
\noindent $\eta^{\mu\tau}\eta^{\nu\sigma} / |z-w|^4$, and is invariant
under separate Lorentz transformations in the indices $\mu, \tau$ and in
the indices $\nu,\sigma$.}
Thus the net symmetry of the coefficients $\tilde A_{\alpha_1\ldots
\alpha_n}$ is $O(d)\times O(d)$ (or $O(d-1,1)\times O(d-1,1)$ if one of
the $d$ coordinates $X^{D-d+1},\ldots X^D$ is time-like)\RSEN.
The same result holds for $\l\tilde\Chi_\alpha|Q_B
c_0^-|\tilde\Chi_\beta\r$.
Of these, the diagonal $O(d)$ subgroup is the usual rotational symmetry of
the original action.
The other generators of the $O(d)\times O(d)$ transformation represent
symmetries of the action only for the restricted set of field
configurations, and hence may be used to generate new classical solutions
of string theory starting from known ones.

To see explicitly how the $O(d)\times O(d)$ transformation acts on the
component fields, let us consider a component field expansion of the form:
$$
|\tilde\Chi\r = \int d^{D-d}k \Big[\big(h_{\gamma\delta}(k)
+b_{\gamma\delta}(k)\big) \alpha^\gamma_{-1}\bar\alpha^\delta_{-1} c_1\bar
c_1  +\phi(k) (c_1c_{-1} -\bar c_1\bar c_{-1})\Big]|k^\gamma=0, k^i\r
+\ldots
\eqn\esftfive
$$
where $X^\gamma$ ($D-d+1\le\gamma\le D$) denote the set of coordinates on
which the solution does not depend, $X^i$ ($1\le i\le D-d$) denote the set
of coordinates on which the solution does depend, $\alpha^\gamma_{-1}$,
$\bar\alpha^\gamma_{-1}$ are the oscillators appearing in the expansion of
the two dimensional fields $X^\gamma$, and $\ldots$ denote (infinite
number of) other terms in the expansion of $|\Chi\r$ that we have not
written down.
Since the $\odd$  transformation corresponds to independent rotation of
holomorphic and anti-holomorphic indices, its action on the fields $h$,
$b$ and $\phi$ is given by,
$$
(h+b)\to S (h+b) R^T, ~~~~ \phi\to\phi
\eqn\esftsix
$$
where $S$ and $R$ are $O(d)$ matrices.
The transformations for which $S=R$ correspond to the usual rotation and
do not generate
inequivalent solutions.
Thus all inequivalent solutions may be generated by taking $S=I$, $R$
arbitrary.

The analysis for heterotic and superstring field theories may be carried
out in the same way.
Since most of our applications will deal with solutions in heterotic
string theory, let us discuss this case briefly.
In this theory, at least in the sector involving the bosonic fields
(Neveu-Schwarz states) the string field theory action may be written in
the same form as in eqs.\esftone,\esfttwoa\RSAROJA, with the difference
that the definition of $A_{r_1\ldots r_n}$ also involves some insertion of
picture changing operators\RFMS.
Thus the same set of arguments show that if we restrict to field
configurations independent of $d$ of the dimensions, the resulting action
will have an $\odd$ symmetry.
In this case, however, the symmetry is even larger.
To see this, let us note that besides the usual bosonic coordinates,
heterotic string theory also has 16 extra bosonic coordinates, which are
purely antiholomorphic functions of the two dimensional coordinates.
Let us denote these by $Y^I$, and consider backgrounds where the
corresponding state $|\tilde\Chi\r$ carry zero momenta conjugate to $p$ of
these $16$ coordinates,\foot{This corresponds to background field
configurations which are invariant
under a $U(1)^p$ subgroup of the gauge group.} besides carrying zero
momentum conjugate to the
$d$ usual bosonic coordinates.
Since the correlation function of such states in the two dimensional
conformal field theory on the plane is invariant under separate rotation
among the $d$ holomorphic and $d+p$ antiholomorphic coordinates, the
resulting action will be invariant under $O(d)\times O(d+p)$
transformation. (Although the coordinates $Y^I$ are compact, the
correlation functions of the kind we are considering are completely
insensitive to this fact, and there is full $O(d+p)$ symmetry involving
rotations among the anti-holomorphic components of $X^\gamma$ and $Y^I$
coordinates.)
As before, the diagonal $O(d)$ subgroup generates usual space-time
rotation; the rest of the generators can be used to generate new solutions
from known ones.

Again to see an example of how this transformation acts on the component
fields, let us consider the following component field expansion of the
heterotic string field in the Neveu-Schwarz sector:
$$\eqalign{
|\Chi\r = & \int d^{D-d}k\Big[ \big( h_{\gamma\delta}(k)
+b_{\gamma\delta}(k)\big )\alpha^\gamma_{-1}\bar\alpha^\delta_{-1} c_1\bar
c_1 +\phi(k) (c_1 c_{-1}-\bar c_1\bar c_{-1}) \cr
&a^I_\gamma(k)\alpha^\gamma_{-1}\bar\beta^I_{-1} c_1\bar c_1 +\ldots
\Big] e^{-\phi}(0)|k^\gamma=0, k^i\r\cr
}
\eqn\esftseven
$$
where $\bar\beta^I_n$ denotes the oscillator associated with the internal
coordinates $Y^I$, and $\phi$ is the bosonized ghost field\RFMS.
The $O(d)\times O(d+p)$ transformation then acts on the oscillators as,
$$\eqalign{
\alpha^\gamma_{-1} &\to S_{\gamma'\gamma}\alpha^{\gamma'}_{-1}\cr
\pmatrix{\bar\alpha^\gamma_{-1}\cr \bar\beta^I_{-1}\cr}&\to R^T
\pmatrix{\bar\alpha^\gamma_{-1}\cr \bar\beta^I_{-1}\cr}
}
\eqn\esfteight
$$
where $S$ and $R$ are $O(d)$ and $O(d+p)$ matrices respectively.
The action of these transformations on the fields $h$, $b$ and $a$ can
easily be seen to be of the form:
$$
(h+b \qquad a) \to S (h+b \qquad a) R^T
\eqn\esftnine
$$
where $(h+b \qquad a)$ is regarded as a $d\times (d+p)$ matrix.

Sometimes we may start with a solution that is invariant under one or more
space-time supersymmetry transformations.
Thus it is natural to ask if the transformed solution is also invariant
under such a symmetry.
A general string field theoretic argument showing that
this must be so may be given as follows.
Since so far we do not have a consistent closed heterotic string field
theory (although we do have such a theory involving only the Neveu-Schwarz
states\RSAROJA) we do not know precisely how the space-time supersymmetry
operator
will look like in the string field theory.
However, from the general analysis\RFMS\ it is clear that space-time
supersymmetry will act only on the holomorphic part of the vertex
operators representing a general off-shell string field configuration.
Since according to the arguments given above the
$O(d+p-1,1)$  symmetry transformation induced by the matrix $R$ acts on
the anti-holomorphic part of the vertex operators, these
two symmetry transformations commute.
On the other hand, the $O(d-1,1)$ transformation induced by the matrix $S$
can
be regarded as a combination of usual Lorentz transformation, and an
$O(d+p-1,1)$ transformation.
Thus the effect of an $S$ transformation will be a usual Lorentz
transformation of the supersymmetry transformation parameter.
As a result, the $O(d-1,1)\times O(d+p-1,1)$ transformation
will transform a
supersymmetric solution to a supersymmetric solution, with the
supersymmetry transformation parameter being Lorentz rotated by the
$O(d-1,1)$
component.

Before I conclude this section, I wish to emphasise that these
results are  exact for the tree level string field
theory,
and no $\alpha'$ expansion has been made.

\chapter{Effective Field Theory Analysis}

{}From the analysis of the last section it follows that the low energy
effective field theory describing heterotic string theory should also
possess the $O(d)\times O(d+p)$ symmetry if we restrict our field
configuration to backgrounds which are independent of $d$ of the $D$
coordinates and is invariant under a $U(1)^p$ subgroup of the gauge group.
In fact if we knew the exact relationship between the component fields
that appear in the expansion of the string field, and those that appear in
the low energy effective field theory, we could find out how the
$O(d)\times O(d+p)$ transformation acts on the fields appearing in the
effective field theory, since we already know how it acts on the component
fields of string field theory.
Unfortunately, the exact relationship between the two sets of fields is
not known, the only known result is in the weak field limit.
Using this, we may express the metric $G_{\mu\nu}$, the dilaton $\Phi$,
the antisymmetric tensor field $B_{\mu\nu}$, and the gauge field $A_\mu^I$
appearing in the effective
field theory in terms of string field components as,
$$\eqalign{
G_{\mu\nu} = & \eta_{\mu\nu} + h_{\mu\nu} + \co(\Chi^2)\cr
B_{\mu\nu} =& b_{\mu\nu} +\co(\Chi^2)\cr
\Phi =& \phi + {1\over 2}  h_{\mu\nu}\eta^{\mu\nu} +\co(\Chi^2)\cr
A_\mu^I =& a_\mu^I +\co(\Chi^2)\cr
}
\eqn\eeffone
$$

Thus the transformation laws of the fields $G_{\mu\nu}$, $B_{\mu\nu}$,
$A_\mu^I$ and $\Phi$ are known explicitly in the weak field limit.
The full $O(d+p)\times O(d)$ transformation laws of these fields are found
by explicitly examining the low energy effective action of the heterotic
string theory.
This action is given by,
$$\eqalign{
S= & -\int d^Dx\sqrt{-\det G}e^{-\Phi}\Big[-R^{(D)}(G)+{1\over 12}
H_{\mu\nu\rho} H^{\mu\nu\rho}\cr
& -G^{\mu\nu}\p_\mu\Phi\p_\nu\Phi
+{1\over 8} \sum_{I=1}^p F^I_{\mu\nu} F^{I\mu\nu} +\ldots \Big]\cr
}
\eqn\eefftwo
$$
where $\ldots$ denote terms in the effective action involving higher
derivative terms, and other fields which are set to zero for the
particular class of backgrounds we are considering, $F^I_{\mu\nu} = \p_\mu
A^I_\nu -\p_\nu A^I_\mu$,
$H_{\mu\nu\rho}=\p_\mu B_{\nu\rho}$ + cyclic permutations $ - (
\Omega^{(3)}(A))_{\mu\nu\rho}$, $R^{(D)}$
denotes the $D$ dimensional Ricci scalar, and $\Omega_3(A)$ is the
Chern-Simons 3-form, defined by,
$$
\Omega_3(A) = {1\over 4} (A^I_\mu F^I_{\nu\rho} + ~cyclic ~permutations)
\eqn\effthree
$$
In writing down the action \eefftwo\ we have restricted ourselves to field
configurations where the background gauge fields are fully abelian,
$p$ denotes the number of unbroken U(1) gauge groups after
compactification.
We shall further restrict to field configurations where the background
gauge fields are independent of $d$ of the $D$ coordinates.
For such restricted set of field configurations, let us define the matrix:
$$
\ck_{\mu\nu} = -G_{\mu\nu} -B_{\mu\nu} -{1\over 4} A^I_\mu A^I_\nu
\eqn\efffour
$$
and treat $G_{\mu\nu}$ and $A^I_\mu$ as $D\times D$ and $D\times p$
matrices respectively.
In terms of these matrices we define a $(2D+p)\times (2D+p)$ matrix as,
$$
\cm=\pmatrix{ (\ck^T-\eta) G^{-1} (\ck -\eta) & (\ck^T-\eta) G^{-1}
(\ck+\eta) & -(\ck^T -\eta) G^{-1} A\cr
(\ck^T+\eta) G^{-1} (\ck -\eta) & (\ck^T+\eta) G^{-1}
(\ck+\eta) & -(\ck^T +\eta) G^{-1} A\cr
-A^T G^{-1} (\ck-\eta) & -A^T G^{-1} (\ck+\eta) & A^T G^{-1} A\cr
}
\eqn\efffive
$$
where $^T$ denotes transposition of the matrix, and $\eta$ is the $D$
dimensional Minkowski metric.
It is easy to see that for a given $\cm$, the fields $G_{\mu\nu}$,
$B_{\mu\nu}$ and $A_\mu^I$ are completely determined.
If we assume that the background field configuration is restricted to be
independent of the last $d$ of the $D$ coordinates $X^{D-d+1}, \ldots
X^D$, then the action for
this restricted field configuration can be shown to be invariant under the
following transformation of the fields\RFAWAD:
$$
\cm\to\cm' = \Omega \cm \Omega^T, ~~~
\Phi\to\Phi' = \Phi + {1\over 2}(\ln\det G' -\ln\det G)
\eqn\eeffsix
$$
where,
$$
\Omega =\pmatrix{I_{D-d} &&&\cr &S&&\cr &&I_{D-d}&\cr &&&R\cr}
\eqn\eeffseven
$$
$S$ and $R$ are $O(d)$ and $O(d+p)$ matrices respectively, and $I_n$ is
the $n$ dimensional identity matrix.
(If the $d$ coordinates on which the solution does not depend includes the
time coordinate, then $S$
and $R$ are $O(d-1,1)$ and $O(d+p-1,1)$ matrices respectively.)
Note that $\cm$ contains full information about the fields $G_{\mu\nu}$,
$B_{\mu\nu}$ and $A^I_\mu$; hence specifying the transformation laws of
$\cm$ and $\Phi$ automatically specifies the transformation laws of all
the fields in the theory.
In order to see that this transformation is the same as the
$O(d)\times
O(d+p)$ transformation discussed in the previous section, we can verify
that they agree with the transformation laws derived in the last section
in the weak field limit.

Using the transformation laws given in eqs.\eeffsix, we can generate new
solutions of the equations of motion from the known solutions that are
independent of some of the coordinates.
In the next two sections we shall see some specific applications of this
procedure of generating new solutions of the equations of motion.

\chapter{Rotating Charged Black Holes in Four Dimensions}

In order to apply the method outlined in the last section for generating
new solutions of the equations of motion, we need to have at least one
known solution of these equations.
As we shall now see, such solutions are provided by already known
solutions of Einstein's equation in the absence of any matter field.
To see this let us
write down the equations of motion following
from the
action given in eq.\eefftwo.
They are,
$$\eqalign{
& R_{\mu\nu} -{1\over 2} G_{\mu\nu}R +D_\mu D_\nu\Phi -G_{\mu\nu}D^\rho
D_\rho \Phi +{1\over 2} G_{\mu\nu}D_\rho\Phi D^\rho\Phi\cr
& -{1\over 4}(H_{\mu\rho\tau} H_{\nu}^{~~\rho\tau} -{1\over 16}
H_{\rho\sigma\tau} H^{\rho\sigma\tau} G_{\mu\nu})
-{1\over 4} (F^I_{\mu\rho} F_\nu^{I~\rho} -{1\over 4} G_{\mu\nu}
F^I_{\rho\tau} F^{I\rho\tau})\cr
=& 0\cr
}
\eqn\eblackone
$$
$$
D_\mu (e^{-\Phi}H^{\mu\nu\rho}) =0
\eqn\eblacktwo
$$
$$
D_\mu(e^{-\Phi} F^{I\mu\nu})+{1\over 2} e^{-\Phi}H_{\rho\mu}^{~~~\nu}
F^{I\rho\mu} =0
\eqn\eblackthree
$$
$$
R - D_\mu\Phi D^\mu\Phi -{1\over 12} H_{\rho\sigma\tau} H^{\rho\sigma\tau}
-{1\over 8} F^I_{\rho\tau} F^{I\rho\tau} +2D^\mu D_\mu\Phi =0
\eqn\eblackfour
$$
{}From eqs.\eblackone-\eblackfour\ we see that if we set
$$
A^I_\mu=B_{\mu\nu} =\Phi =0
\eqn\eblackfive
$$
then all the equations are satisfied if the
metric satisfies the Einstein's equation $R_{\mu\nu} -{1\over 2}
G_{\mu\nu} R=0$.
In other words, given a solution of the pure Einstein's equation, we can
construct a solution of the full string theory equations of motion by
setting $A^I_\mu=B_{\mu\nu} =\Phi =0$.

We now restrict ourselves to four dimensions, and consider the most
general static black hole solution of Einstein's equation.
This is given by the Kerr solution:
$$\eqalign{
ds^2 =& -{\rho^2 +a^2\cos^2\theta -2m\rho\over \rho^2
+a^2\cos^2\theta} dt^2 +{\rho^2 +a^2\cos^2\theta\over\rho^2 +a^2 -2m\rho}
d\rho^2 + (\rho^2 +a^2\cos^2\theta) d\theta^2\cr
& + {\sin^2\theta\over \rho^2 + a^2\cos^2\theta} \{(\rho^2 +a^2)(\rho^2
+a^2\cos^2\theta) +2m\rho a^2\sin^2\theta\}d\phi^2\cr
& -{4m\rho a\sin^2\theta\over \rho^2 +a^2\cos^2\theta} dt d\phi \cr
}
\eqn\eblacksix
$$
This metric, together with the field configuration given in
eq.\eblackfive, gives a solution of the equations of motion
\eblackone-\eblackfour.

We now note that the solution is independent of the time coordinate.
Hence we can generate new solutions by making a
transformation of the form given in eqs.\eeffsix, \eeffseven, with,
$$
S=I_p, ~~~~R = \pmatrix{\cosh\alpha & \sinh\alpha & \cr \sinh\alpha &
\cosh\alpha & \cr && I_{p-1}\cr}
\eqn\eblackseven
$$
After some algebraic manipulations, the transformed solution may be
expressed as\RROTBLA,
$$\eqalign{
ds^{\prime 2}=& -{(\rho^2 +a^2\cos^2\theta -2m\rho)
(\rho^2+a^2\cos^2\theta) \over (\rho^2+a^2\cos^2\theta
+2m\rho\sinh^2{\alpha\over 2})^2} dt^2\cr
& +{\rho^2 +a^2\cos^2\theta\over \rho^2
+a^2 -2m\rho} d\rho^2 + (\rho^2 +a^2\cos^2\theta) d\theta^2\cr
&+\{(\rho^2+a^2)(\rho^2+a^2\cos^2\theta) +2m\rho a^2\sin^2\theta +4m\rho
(\rho^2+a^2) \sinh^2{\alpha\over 2}\cr
& + 4m^2\rho^2\sinh^4{\alpha\over 2}\}
 \times {(\rho^2+a^2\cos^2\theta)\sin^2\theta\over (\rho^2
+a^2\cos^2\theta +2m\rho\sinh^2{\alpha\over 2})^2} d\phi^2\cr
&- {4m\rho a\cosh^2{\alpha\over 2} (\rho^2+a^2\cos^2\theta)\sin^2\theta
\over (\rho^2
+a^2\cos^2\theta +2m\rho\sinh^2{\alpha\over 2})^2} dt d\phi\cr
}
\eqn\eten
$$
$$
\Phi' =-\ln {\rho^2 +a^2\cos^2\theta +2m\rho\sinh^2{\alpha\over 2} \over
\rho^2 +a^2\cos^2\theta}
\eqn\eeleven
$$
$$
A'_\phi = -{2m\rho a\sinh\alpha\sin^2\theta\over \rho^2 +a^2\cos^2\theta
+2m\rho\sinh^2{\alpha\over 2}}
\eqn\etwelve
$$
$$
A'_t = {2m\rho\sinh\alpha\over \rho^2 +a^2\cos^2\theta
+2m\rho\sinh^2{\alpha\over 2}}
\eqn\ethirteen
$$
$$
B'_{t\phi} = {2m\rho a\sinh^2{\alpha\over 2}\sin^2\theta \over \rho^2
+a^2\cos^2\theta +2m\rho\sinh^2{\alpha\over 2}}
\eqn\efourteen
$$
The other components of $A'_\mu$ and $B'_{\mu\nu}$ vanish.
(Here $A_\mu$ denotes the component $A_\mu^1$.)
The Einstein metric (defined as $ds^{\prime 2}_E\equiv
e^{-\Phi'}ds^{\prime 2}$) is given
by,
$$\eqalign{
ds^{\prime 2}_E =& -{\rho^2 +a^2\cos^2\theta -2m\rho\over \rho^2
+a^2\cos^2\theta +2m\rho\sinh^2{\alpha\over 2}} dt^2
+{\rho^2
+a^2\cos^2\theta +2m\rho\sinh^2{\alpha\over 2}\over \rho^2 +a^2 -2m\rho}
d\rho^2 \cr
& +(\rho^2
+a^2\cos^2\theta +2m\rho\sinh^2{\alpha\over 2}) d\theta^2
-{4m\rho a\cosh^2{\alpha\over 2}\sin^2\theta\over \rho^2
+a^2\cos^2\theta +2m\rho\sinh^2{\alpha\over 2}} dtd\phi\cr
& +\{(\rho^2+a^2)(\rho^2+a^2\cos^2\theta) +2m\rho a^2\sin^2\theta +4m\rho
(\rho^2+a^2) \sinh^2{\alpha\over 2}\cr
& + 4m^2\rho^2\sinh^4{\alpha\over 2}\}
\times {\sin^2\theta \over \rho^2
+a^2\cos^2\theta +2m\rho\sinh^2{\alpha\over 2}} d\phi^2\cr
}
\eqn\efifteen
$$
The various field strengths associated with the gauge fields are given by,
$$
F_{\rho\phi} ={2ma\sinh\alpha\sin^2\theta (\rho^2-a^2\cos^2\theta) \over
(\rho^2 +a^2\cos^2\theta +2m\rho\sinh^2{\alpha\over 2})^2}
\eqn\ethreesix
$$
$$
F_{\theta\phi} =- {4m\rho a\sinh\alpha (\rho^2 +a^2
+2m\rho\sinh^2{\alpha\over 2})\sin\theta\cos\theta \over \den^2}
\eqn\ethreeseven
$$
$$
F_{\rho t} = -{2m\sinh\alpha (\rho^2 -a^2\cos^2\theta)\over \den^2}
\eqn\ethreeeight
$$
$$
F_{\theta t} ={4m\rho a^2\sinh\alpha\sin\theta\cos\theta \over\den^2}
\eqn\ethreenine
$$
$$
e^{-\Phi}\sqrt{-G}H^{\rho t\phi} ={2ma\sinh^2{\alpha\over 2} (\rho^2 -
a^2\cos^2\theta)\sin\theta\over (\rho^2 +a^2\cos^2\theta)^2}
\eqn\ethreeten
$$
$$
e^{-\Phi}\sqrt{-G} H^{\theta\phi t} ={4m\rho a\sinh^2{\alpha\over
2}\cos\theta \over (\rho^2 +a^2\cos^2\theta)^2}
\eqn\ethreeeleven
$$

By examining the asymptotic properties of the solution we can easily
determine that this solution describes a rotating object with mass $M$,
angular momentum $J$, charge $Q$ and magnetic moment $\mu$, given
by,\foot{A factor of $2\sqrt 2$ in the definition of the electric charge and
electric current has been introduced so that our normalization matches
that of ref.\RHOST.}
$$\eqalign{
M =& {m\over 2} (1+\cosh\alpha), ~~~~ Q ={m\over\sqrt 2} \sinh\alpha\cr
J =& {ma\over 2} (1+\cosh\alpha), ~~~~ \mu ={1\over\sqrt 2} ma\sinh\alpha
}
\eqn\eblackeight
$$
Hence the $g$ factor is given by,
$$
g\equiv {2\mu M\over QJ} = 2
\eqn\eeighteen
$$
For $a=0$, the above solution reduces to the charged black hole
solution of refs.\RGIB\RHOST.
Construction of rotating charged black hole solution in string theory in
the limit of small angular momentum had been discussed previously by Horne
and Horowitz\RHOHO.

By examining the above solution, we see that it has two horizons at,
$$
\rho =m \pm \sqrt{m^2 - a^2} = M - {Q^2\over 2M} \pm \sqrt{(M-{Q^2\over
2M})^2 -{J^2\over M^2}}
\eqn\eblacknine
$$
These event horizons exist as long as,
$$
M-{Q^2\over 2M}\ge {|J|\over M}
\eqn\eblackten
$$
Thus the extremal limit, the limit in which the event horizon is about to
disappear, is given by,
$$
M^2\to {Q^2\over 2} +|J|
\eqn\eblackeleven
$$
The area of the
outer event horizon, which is proportional to the entropy of the black
hole, is given by,
$$
A= 8\pi M\bigg(M -{Q^2\over 2M} +\sqrt{(M-{Q^2\over 2M})^2 -{J^2\over
M^2}}\bigg)
\eqn\eblacktwelve
$$
Thus in the extremal limit it approaches the value $8\pi |J|$.

Finally, the surface gravity of the black hole, which is proportional to
the Hawking temperature of the black hole, is given by,
$$
{\sqrt{(2M^2 - Q^2)^2 -4J^2}\over 2M(2M^2 - Q^2 +\sqrt{(2M^2 - Q^2)^2 -
4J^2})}
\eqn\eblackthirteen
$$
Note that the surface gravity vanishes in the extremal limit.

\chapter{Macroscopic Charged Heterotic String}

We now turn to the second example.
In this case we start from a solution of the string theory equations of
motion given in ref.\RDGHR, describing fundamental heterotic string in
$D$ dimensions.
The solution is given by,
$$\eqalign{
ds^2=&e^E\{ -dt^2 +(dx^{D-1})^2\} +\sum_{i=1}^{D-2} dx^i dx^i\cr
B_{(D-1)t}=&1-e^E,~~~\Phi=E,~~~ A_\mu^{I}=0\cr
}
\eqn\etwoeighta
$$
where,
$$
e^{-E}=1 +MG^{(D-2)}(\vec r)
\eqn\etwoeightb
$$
$\vec r$ denotes the $D-2$ dimensional vector $(x^1,\ldots x^{D-2})$, and
$G^{(D-2)}$ is the $D-2$ dimensional Green's function, given by,
$$\eqalign{
G^{(D-2)}(\vec r)=& {1\over (D-4)\omega_{D-3} r^{D-4}}~~~{\rm for}~D>4\cr
=& -{1\over 2\pi}\ln r~~~{\rm for}~D=4\cr
}
\eqn\etwoeightc
$$
where $\omega_{D-3}$ is the volume of a unit $D-3$ sphere.
$M$ is an arbitrary constant.
By looking at the source terms at $\vec r=0$ that this solution
corresponds to, we can identify the solution to be the field around a
fundamental heterotic string\RDGHR\ sitting at $\vec r=0$ in the gauge
$X^0=\tau$, $X^{D-1}=\sigma$, $(\sqrt{-\gamma})^{-1}\gamma_{mn}
=\eta_{mn}$, provided we identify
the constant $M$ appearing in the solution to the string tension.
(Here $\gamma_{mn}$ denotes the world sheet metric).

By examining the solution we see that it is independent of the coordinates
$t$ and $x^{D-1}$.
Hence we can generate new solutions by performing transformations of the
kind discussed in sec.3.
We choose a transformation matrix $\Omega$ of the form:
$$
\Omega =\pmatrix{I_{2D-1} &&&\cr & \cosh\alpha &\sinh\alpha &\cr &
\sinh\alpha &\cosh\alpha &\cr &&& I_{p-1}\cr }
\eqn\etwothirteen
$$
Using eq.\eeffsix, and some algebra, we find the
transformed solution to be\RSTRING:
$$\eqalign{
ds^2 =& {1\over 1+NG^{(D-2)}(\vec r)} (-dt^2 +(dx^{D-1})^2)
+{q^2 G^{(D-2)}(\vec r)\over 4N ( 1+NG^{(D-2)}(\vec r))^2}
(dt+dx^{D-1})^2\cr
&+\sum_{i=1}^{D-2} dx^i dx^i\cr\cr
B_{(D-1)t} =& {NG^{(D-2)}(\vec r)\over  1+NG^{(D-2)}(\vec r)}\cr\cr
A^{1}_{D-1} =& A^{1}_t = {q G^{(D-2)}(\vec r)\over  1+NG^{(D-2)}(\vec
r) }\cr\cr
\Phi= &-\ln( 1+NG^{(D-2)}(\vec r))\cr\cr
}
\eqn\etwofifteen
$$
where
$$
N=M\cosh^2{\alpha\over 2},~~~~~q=M\sinh\alpha
\eqn\etwosixteen
$$
The various field strength tensors may be calculated from this solution
and we get the following results,
$$\eqalign{
F^{1}_{r (D-1)} =& F^{1}_{rt} = - {q\over r^{D-3}\omega_{D-3} (
1+NG^{(D-2)}(\vec r))^2}\cr
H_{r (D-1) t} =& -{N\over  r^{D-3}\omega_{D-3} (
1+NG^{(D-2)}(\vec r))^2}\cr
}
\eqn\etwoseventeen
$$
Note that $\Omega_3(A)$ vanishes everywhere for the specific solution we
have constructed.
For $D>4$ the metric is asymptotically
flat, and the electric charge per unit length $Q$, the electric current
$J$ and the
axionic charge
$Z$ associated with the solution may be defined in terms of the asymptotic
behavior of the field strengths in the $r\to\infty$ limit as follows:
$$\eqalign{
{1\over 2\sqrt 2} F^{1}_{rt} &\simeq -{Q\over r^{D-3}\omega_{D-3}}\cr
{1\over 2\sqrt 2} F^{1}_{r (D-1)} &\simeq -{J\over
r^{D-3}\omega_{D-3}}\cr
H_{r (D-1) t} &\simeq -{Z\over r^{D-3}\omega_{D-3}}\cr
}
\eqn\etwoeighteen
$$
Eqs.\etwoseventeen\ and \etwoeighteen\ give,
$$
Q={q\over 2\sqrt 2}, ~~~~ J={q\over 2\sqrt 2}, ~~~~ Z=N
\eqn\etwonineteen
$$
Note that the gauge field configurations associated with the solution
describe a radial electric field and an azimuthal magnetic field, which
are equal in magnitude.
This, in turn, shows that the solution corresponds to a charge and current
carrying string, for which the electric current is equal to the charge per
unit length of the string.

Analysis of the source terms associated with the solution near the origin
$\vec r=0$ reveals that the new solution also describes a fundamental
string with string tension $N$, carrying a world sheet current $j^m$
satisfying $j^0=-j^1$.
This is in accordance with the well known property of the heterotic string
that the world-sheet currents in this theory responsible for the gauge
symmetry are chiral in nature.

Before concluding this section let us mention some special properties of
the solution.
The original solution of ref.\RDGHR\ was shown to be invariant under half
of the space-time supersymmetry transformations of the theory.
The general argument given in section 2 then shows that the transformed
solution should also be invariant under half of the supersymmetry
transformations.
In fact, since
in the present case, the transformation is generated solely by the
$O(d+p-1,1)$ part (we see from eqs.\eeffseven\ and \etwothirteen\ that $S=I$
in the present case), we expect from the general arguments that the
transformed solution will be
invariant under the same supersymmetry transformations as the original
solution.
These results can be verified by explicit calculation.
In fact, recently the charged string solution has been reconstructed by
demanding that the final solution is supersymmetry invariant\RNEWST.

It was shown that the solution of ref.\RDGHR\ can be regarded as the
extremal limit of a black string solution\RHOST.
The present solution can also be shown to be the extremal limit of a
charged black string solution\RSTRING.
One can simply start from the black string solution of ref.\RHOST\ and
transform it by the $O(d-1,1)\times O(d+p-1,1)$ transformation to find out
the
black string solution whose extremal limit is the solution given above.

The authors of ref.\RDGHR\ also constructed static multi-string solution,
representing parallel strings at rest.
An $O(d-1,1)\times O(d+p-1,1)$ transformation their solution gives a
static, charged, multi-string solution.
This, however, is not the most general charged multi-string solution,
since the charge densities carried by all the strings are parametrized by
a single parameter (the boost angle $\alpha$ appearing in
eq.\etwothirteen).
However, a slight modification of the solution thus obtained allows us to
construct charged multi-string solutions, where the charge densities
carried by different strings are independent of each other, and can also
point in arbitrary directions in the $p$ dimensional space corresponding
to the $p$ $U(1)$ gauge groups of the theory.
This general solution is given by\RSTRING,
$$\eqalign{
ds^2 =& {1\over 1+N\sum_l G^{(D-2)}(\vec r-\vec r_l)} (-dt^2
+(dx^{D-1})^2)\cr
& + g(\vec r)
(dt+dx^{D-1})^2
+\sum_{i=1}^{D-2} dx^i dx^i\cr
B_{(D-1)t} =& {N\sum_l G^{(D-2)}(\vec r-\vec r_l)\over
1+N\sum_lG^{(D-2)}(\vec r-\vec r_l)}\cr
A^{I}_{D-1} =& A^{I}_t = { \sum_l q_l^{(I)} G^{(D-2)}(\vec r-\vec
r_l)\over
1+N\sum_l G^{(D-2)}(\vec
r-\vec r_l) }\cr
\Phi= &-\ln( 1+N\sum_l G^{(D-2)}(\vec r-\vec r_l))\cr
}
\eqn\estringthreeseven
$$
where,
$$
g(\vec r) ={1\over 4} \bigg[{\sum_{l,I}(q^{(I)}_l)^2 G^{(D-2)}(\vec r-\vec
r_l)\over N\big(1 +\sum_l NG^{(D-2)}(\vec r-\vec r_l)\big)}
-{\sum_I\big(\sum_l q_l^{(I)} G^{(D-2)}(\vec r-\vec r_l)\big)^2 \over
\big(1 +\sum_l NG^{(D-2)}(\vec r-\vec r_l)\big)^2}\bigg]
\eqn\estringthreenine
$$

\chapter{Black Holes Carrying Electric and Magnetic Charges}

Besides the $O(d)\times O(d+p)$ invariance for restricted class of
backgrounds of the form discussed above, in four dimensions the equations
of motion derived from the action \eefftwo\ have another symmetry, known
as the electric-magnetic duality symmetry\RGAZU,\RODD.
Using this symmetry one can generate solutions carrying both electric and
magnetic charges starting from solutions carrying only electric
charge\RTSW\RSDUAL\RORTIN.
Unlike the $O(d)\times O(d+p)$ symmetry discussed in the previous
sections, the transformation laws under this symmetry take simple form in
terms of the Einstein metric $G_{E\mu\nu}= e^{-\Phi}G_{\mu\nu}$.
For convenience of writing, throughout this section we shall use the
notation $G_{\mu\nu}$ for Einstein metric, and all the indices will be
raised and lowered with this metric rather than the $\sigma$-model metric
used in the last sections.

In order to understand the action of this symmetry, we first note that
for any solution of the equations of motion, eq.\eblacktwo\ allows us to
define a field $\Psi$  as follows:
$$
H^{\mu\nu\rho} =-(\sqrt{-\det G})^{-1} e^{2\Phi}\eps^{\mu\nu\rho\sigma}
\p_\sigma \Psi
\eqn\edualtwoeleven
$$
(Note that the $G_{\mu\nu}$ ($H^{\mu\nu\rho}$) appearing in the above
equation is related to
$G_{\mu\nu}$ ($H^{\mu\nu\rho}$)
appearing in eq.\eblacktwo\ by a multiplicative factor of $e^{-\Phi}$
($e^{3\Phi}$); this
is the reason for the appearance of the factor of $e^{2\Phi}$ on the right
hand side of eq.\edualtwoeleven\ instead of $e^\Phi$.)
Let us define,
$$
\tilde F^{I\mu\nu} ={1\over 2} \sqrt{-\det G}\epsilon^{\mu\nu\rho\sigma}
F^I_{\rho\sigma}
\eqn\etttone
$$
The bianchi identity of the field $H_{\mu\nu\rho}$,
$$
(\sqrt{-\det G})^{-1}\eps^{\mu\nu\rho\sigma}\p_\mu H_{\nu\rho\sigma}
=-{3\over  4} F^I_{\mu\nu}\tilde F^{I\mu\nu}
\eqn\edualtwonine
$$
may then be written as,
$$
D^\mu (e^{2\Phi}D_\mu \Psi) ={1\over 8} \sum_{I=1}^p
F^I_{\mu\nu}\tF^{I\mu\nu}
\eqn\edualtwotwelve
$$
Let us now define a complex field $\lambda$ as,
$$
\lambda = \Psi + ie^{-\Phi} \equiv \lambda_1 + i\lambda_2
\eqn\edualtwothirteen
$$
The equations of motion \eblackone, \eblackthree, \eblackfour, and
\edualtwotwelve\ may then be written as,
$$
R_{\mu\nu} ={ \p_\mu\bl \p_\nu\lambda +\p_\nu\bl \p_\mu\lambda \over
4(\lambda_2)^2} + {1\over 4} \lambda_2 F^I_{\mu\rho}F_{\nu}^{I~\rho}
-{1\over 16}\lambda_2 G_{\mu\nu} F^I_{\rho\sigma} F^{I\rho\sigma}
\eqn\edualtwofourteen
$$
$$
{D^\mu D_\mu\lambda \over (\lambda_2)^2} + i {D_\mu\lambda D^\mu\lambda
\over (\lambda_2)^3} -{i\over 16} F^I_{-\mu\nu}F_-^{I~\mu\nu} =0
\eqn\edualtwofifteen
$$
$$
D_\mu (\lambda F_+^{I~\mu\nu} -\bar\lambda F_-^{I~\mu\nu}) =0
\eqn\edualtwosixteen
$$
where,
$$
F^I_{\pm} = F^I\pm i\tF^I
\eqn\edualtwoseventeen
$$
In terms of the fields $F^I_\pm$, the Bianchi identity for $F^I_{\mu\nu}$
takes the form:
$$
D_\mu (F_+^{I~\mu\nu} - F_-^{I~\mu\nu}) =0
\eqn\edualtwoeighteen
$$
We now note that eqs.\edualtwofourteen-\edualtwosixteen, and
\edualtwoeighteen\ are
invariant under the following transformations:
$$
\lambda \to \lambda + c
\eqn\edualtwoeighteena
$$
where $c$ is a real number, and,
$$
\lambda \to -{1\over\lambda}, ~~~~ F^I_+\to -\lambda F^I_+, ~~~~ F^I_-\to
-\bl F^I_-
\eqn\edualtwoeighteenb
$$
Invariance of all the equations under \edualtwoeighteena\ is manifest.
Under \edualtwoeighteenb, eq.\edualtwofifteen\ is invariant,
eqs.\edualtwosixteen\ and
\edualtwoeighteen\ get interchanged, and eq.\edualtwofourteen\ transforms
to itself plus an extra term, given by,
$$
-{\lambda_1 (\lambda_2)^2\over |\lambda|^2} (2 F^I_{\mu\rho}
\tF_\nu^{I~\rho} + 2 F^I_{\nu\rho} \tF_\mu^{I~\rho} -g_{\mu\nu}
F^I_{\rho\tau} \tF^{I\rho\tau})
\eqn\edualtwonineteen
$$
The term given in eq.\edualtwonineteen, however, vanishes identically in
four dimensions, showing that \edualtwoeighteenb\ is a genuine symmetry of
the equations of motion.
The two transformations together generate the full SL(2,\R) group under
which $\lambda\to (a\lambda +b)/(c\lambda +d)$ with $ad-bc=1$, and
$F^I_+\to -(c\lambda +d) F^I_+$.

Applying this `symmetry' transformation on the known solutions of the
equations of motion, we can find new solutions.
Let us now apply this transformation to the rotating charged black hole
solution given in sect.4.
{}From eqs.\ethreeten, \ethreeeleven, and \edualtwoeleven\ we get,
$$
\Psi={2ma\sinh^2{\alpha\over 2}\cos\theta
 \over \rho^2 + a^2\cos^2\theta}
\eqn\ethreetwelve
$$
(Note that we must multiply $G$ by appropriate factor of $e^\Phi$ before
comparing eqs.\ethreeten, \ethreeeleven\ with \edualtwoeleven.)
This gives,
$$
\lambda = {2ma\sinh^2{\alpha\over 2}\cos\theta +i (\rho^2 +a^2\cos^2\theta
+2m\rho\sinh^2{\alpha\over 2})  \over \rho^2 +
a^2\cos^2\theta}
\eqn\eduallambda
$$
The dual of various field strengths $F_{\mu\nu}$ are given by,
$$
\tF_{\rho\phi}={4m\rho a^2\sinh\alpha\sin^2\theta\cos\theta \over \den^2}
\eqn\edualdualone
$$
$$
\tF_{\theta t} = -{2ma\sinh\alpha(\rho^2- a^2\cos^2\theta)\sin\theta
\over\den^2}
\eqn\edualdualtwo
$$
$$
\tF_{\theta\phi} = {2m\sinh\alpha\sin\theta (\rho^2 - a^2\cos^2\theta)
(\rho^2 + a^2 + 2m\rho\sinh^2{\alpha\over 2})\over\den^2}
\eqn\edualdualthree
$$
$$
\tF_{\rho t} =-{4m\rho a\sinh\alpha \cos\theta \over\den^2}
\eqn\edualdualfour
$$
We can now generate a new solution by performing an SL(2,\R) transformation
on the above solution.
We consider the SL(2,\R) transformation $\lambda\to -(1+c^2)/(\lambda+c)$,
$F_+\to -(\lambda +c)F_+/\sqrt{1 +c^2}$.
The transformed solution is given by,
$$
\lambda ' =-{1+c^2\over \lambda +c}, ~~~~ ds^{\prime 2} = ds_E^{\prime 2},
{}~~~~
F'_{\mu\nu} = - {\Psi+c\over \sqrt{1+c^2}} F_{\mu\nu} +
{e^{-\Phi}\over \sqrt{1+c^2}} \tF_{\mu\nu} \
\eqn\ethreedualthirteen
$$
where $\lambda$, $ds_E^{\prime 2}$, $F_{\mu\nu}$ and $\tF_{\mu\nu}$ are
given in eqs.\eduallambda, \efifteen, \ethreesix-\ethreenine, and
\edualdualone-\edualdualfour\ respectively.
I shall not write out the solution in detail.
The leading components of the electromagnetic field at large $\rho$ are
given by,
$$
F'_{\rho t} \simeq {2mc\sinh\alpha\over \sqrt{1+c^2} \rho^2}, ~~~~
F'_{\theta\phi} = {2m\sinh\alpha\sin\theta\over\sqrt{1+c^2}}
\eqn\edualthreefourteen
$$
With appropriate normalization (which sets the coefficient of $F^2$ term
in the action to unity),  the electric and magnetic charges carried
by the solution may be identified to,
$$
Q_{el}={1\over\sqrt 2} {mc\sinh\alpha\over\sqrt{1+c^2}}, ~~~~
Q_{mag} = {1\over\sqrt 2}{m\sinh\alpha\over\sqrt{1+c^2}}
\eqn\ethreeseventeen
$$
Since the metric remains the same, the expressions for the mass $M$ and
angular momentum $J$ of the black hole in terms of the parameters $m$, $a$
and  $\alpha$ remain the same as in eq.\eblackeight.
Also, all geometric properties of the
solution remain the same as in the case of rotating charged black hole.
The only difference that occurs in eqs.\eblacknine-\eblackthirteen\ is the
replacement of $Q^2$ by $(Q_{el})^2 +(Q_{mag})^2$.
Thus this solution
represents a rotating black hole solution carrying both
electric and magnetic charge\RSDUAL.
To linear order in the electric and magnetic charges, this solution was
previously constructed in ref.\RCAMP.

\chapter{Summary}

In this review I have discussed the construction of various black hole and
soliton type solutions in string theory using a general method,
which relies on the existence of extended symmetries of the equations of
motion when we restrict the field configuration to a certain class. Two
such extended symmetries have been discussed, one of which appears when we
restrict field configurations to be independent of certain directions, and
no charged fields are present; the other appears when we restrict the field
configurations to four dimensions, and again no charged fields are present.
Combining these two symmetry transformations we can generate solutions
carrying both electric and magnetic charges by starting from a solution of
vacuum Einstein's equation.
The method is quite powerful, and using this method one can
also generate  most of the known black
hole type solutions in critical string theory (where the dilaton
approaches a constant value asymptotically)  by starting
from a known solution of vacuum Einstein's equation in some dimension and
then repeatedly applying these symmetry transformations.

\refout
\end